\pdfoutput=1
\documentclass[aps,prl,floatfix,twocolumn]{revtex4-2}
\usepackage{graphicx}
\usepackage{amssymb}
\usepackage[squaren, cdot]{SIunits}
\usepackage[latin9]{inputenc}
\usepackage{amsmath}
\usepackage{lineno}
\usepackage{changes}
\usepackage{hyperref}

\newcommand*{\figref}[2][]{%
	\hyperref[{fig:#2}]{%
		Fig.~\ref*{fig:#2}%
		\ifx\\#1\\%
		\else
		\,(#1)%
		\fi
	}%
}

\newcommand*{\figureref}[2][]{%
	\hyperref[{fig:#2}]{%
		Figure~\ref*{fig:#2}%
		\ifx\\#1\\%
		\else
		\,(#1)%
		\fi
	}%
}

\begin{document}
\title{Defect-Free Assembly of 2D Clusters of More Than\\
	100 Single-Atom Quantum Systems}
\author{Daniel Ohl de Mello}
\author{Dominik Sch\"{a}ffner}
\author{Jan Werkmann}
\author{Tilman Preuschoff}
\author{Lars Kohfahl}
\author{Malte Schlosser}
\author{Gerhard Birkl}
\email[]{gerhard.birkl@physik.tu-darmstadt.de}
\homepage{www.iap.tu-darmstadt.de/apq}
\affiliation{Institut f\"{u}r Angewandte Physik, Technische Universit\"{a}t Darmstadt, Schlossgartenstra\ss e 7, 64289 Darmstadt, Germany}
\date{\today}

\begin{abstract}
	We demonstrate the defect-free assembly of versatile target patterns of up 111 neutral atoms, building on a 361-site subset of a micro-optical architecture that readily provides thousands of sites for single-atom quantum systems.
	By performing multiple assembly cycles in rapid succession, we drastically increase achievable structure sizes and success probabilities. 
	We implement repeated target pattern reconstruction after atom loss and deterministic transport of partial atom clusters necessary for distributing entanglement in large-scale systems.
	This technique will propel assembled-atom architectures beyond the threshold of quantum advantage and into a regime with abundant applications in quantum sensing and metrology, Rydberg-state mediated quantum simulation, and error-corrected quantum computation.\\
	
	\doi{10.1103/PhysRevLett.122.203601}
\end{abstract}

\maketitle

The next major breakthrough in quantum science and technology necessitates experimental platforms that provide extensive scalability, multisite quantum correlations, and efficient quantum error correction \cite{Acin2018}. Formidable progress has been reported for various systems. Among them, neutral atoms trapped by light are of specific interest since they
offer well-isolated quantum systems with favorable scaling properties \cite{Schlosser2011,Piotrowicz2013,Endres2016,Barredo2016,Kim2016,Gross2017,Barredo2018,Kumar2018,Brown2019}, comprehensive quantum-state control, and on-demand interaction processes \cite{SaffmanWalkerMolmer2010,Saffman2016,Browaeys2016}.
Further progress is crucially dependent on the reliable realization of defect-free target structures.
For the submicron spaced periodic potentials of optical lattices, the preparation of a central region with near-unity filling has been demonstrated in two-dimensional (2D) quantum gas microscopes \cite{Gross2017,Bakr2010,Sherson2010,Zeiher2015,Mazurenko2017,Chiu2018,Guardado2018}. Accurate repositioning of individual atoms has been implemented for four atoms in a one-dimensional polarization-synthesized optical lattice \cite{Robens2017}, but unrestricted individual atom transport remains a challenge in these lattices for higher atom numbers and dimensionality.
\\ \indent
In focused beam microtrap arrays with spacings in the micrometer regime, individual atoms are prepared directly from a thermal ensemble through collisional blockade \cite{SchlosserN2001a,Grunzweig2010,Schlosser2012,Lester2015}. Efficiencies can reach \unit{90}{\%}, as demonstrated for up to four traps \cite{Brown2019,Grunzweig2010,Lester2015}, but are typically about \unit{50}{\%} for larger systems. Thus, additional atom rearrangement is required to eliminate defects.
A 51-atom quantum simulator has been demonstrated based on a linear optical tweezer array generated by a multitone acousto-optic deflector (AOD) and atom-sorting through muting unoccupied sites and compressing the occupied ones \cite{Endres2016,Brown2019,Bernien2017}.
A different approach is based on configuring a desired light field by the use of a 2D liquid-crystal spatial light modulator (SLM) \cite{Barredo2016,Kim2016,Barredo2018}.
This leads to holographically created trap arrays with adaptable geometries. Atom relocation has been demonstrated either by rearrangement of the traps themselves through the sequential altering of the pixel-based phase pattern \cite{Kim2016} or by using a superposed moving optical tweezer \cite{Barredo2016}. 
In these systems, the simulation of spin Hamiltonians \cite{Kim2018} and the realization of topologically protected bosonic phases \cite{Leseleuc2018} have been achieved. The extension of this approach to pattern formation in three dimensions with up to 72 atoms \cite{Barredo2018} and
the application of a large-spacing three-dimensional optical lattice for
the realization of Maxwell's demon with 60 atoms \cite{Kumar2018} have been reported recently.
\\ \indent
All prospect applications of assembled-atom platforms in quantum science and technology will strongly benefit from scaling the system size to larger atom numbers. This crucially depends on the initial number of source atoms, the success probability of target structure assembly, and the ability to mend atom loss during operation. In this Letter, we introduce a unique micro-optical platform for neutral-atom-based quantum engineering that does not experience the limiting effects of size restrictions due to the finite frequency spectrum of AODs and constraints in pixelation and laser power resistance of SLMs. In addition to its outstanding scaling properties with the near-term prospect of incorporating thousands of individually addressable sites, our platform gives access to scaling up the size of the achievable atom array, through the efficient utilization of a large reservoir of atoms in consecutive assembly cycles, to stabilization and reconstruction of target structures and to multiple repetitions of quantum algorithms within a single cooling and trapping cycle. With trap separations in the range of micrometers, our platform is well suited for the implementation of Rydberg-state mediated interactions \cite{SaffmanWalkerMolmer2010,Saffman2016,Browaeys2016} for quantum simulation and computation.
\begin{figure*}[htb]
	\includegraphics[width=\linewidth]{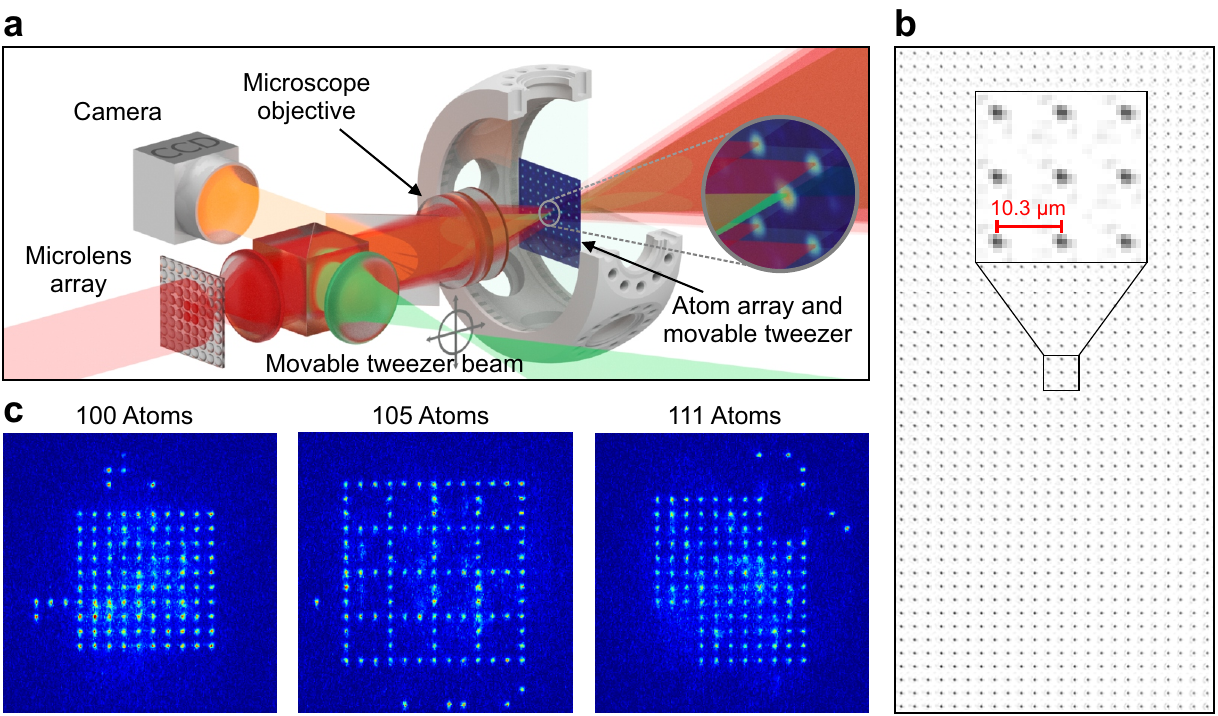}
	\caption{Demonstration of defect-free $N \geq 100$ atom clusters and experimental setup. (a) Simplified schematic of the experimental setup. 
	(b) Reimaged focal plane of the microlens array. Here, a 1200-site subregion out of the total array containing more than 2500 focal spots is shown. We observe excellent homogeneity of waists [$\unit{1.45(10)}{\micro\metre}$] and pitch [$\unit{10.3(3)}{\micro\metre}$] over the whole array.  
	(c)	Various defect-free clusters with 100, 105, and 111 atoms, respectively.}
	\label{fig:setup}
\end{figure*}
\\ \indent
As depicted in \figref[a]{setup}, we create an array of focused-beam dipole traps with tunable separations from the focal spot pattern of a microlens array (MLA) reimaged into a vacuum chamber.
This directly links each atom trap to a specific illuminated microlens, significantly reducing complexity and ensuring laser power efficiency and scalability. Our approach is readily capable of providing thousands of microtraps in a 2D plane [see \figref[b]{setup}], and we create a variety of compact, defect-free clusters of up to 111 atoms [\figref[c]{setup}]. Using a MLA with a pitch of $\unit{110}{\micro\metre}$ and a microscope objective with an effective focal length of $\unit{37.5(10)}{\milli\metre}$ and a numerical aperture of $0.25(2)$, the
experiments reported here are performed in a $\unit{10.3(3)}{\micro\metre}$-pitch quadratic array of traps with beam waists of $\unit{1.45(10)}{\micro\metre}$. We utilize a workspace of 361 sites in a $19 \times 19$ grid.
The trapping light wavelength is $\unit{797.3}{\nano\metre}$ and for rubidium atoms the trap depths are $U_0/k_B = \unit{0.21(3)\,...\,1.7(2)}{\milli\kelvin}$ (grid corner to center), due to the Gaussian profile of the beam illuminating the MLA. For atom transport, we superpose a moving optical tweezer steered by a 2D AOD, which is slightly offset in wavelength to avoid interference effects. Its focal waist is $\unit{2.0(1)}{\micro\meter}$, corresponding to a trap depth of $U_0^{'}/k_B = \unit{0.52(5)}{\milli\kelvin}$. The addressable region encloses more than 1500 sites.
Starting from a magneto-optical trap followed by an optical molasses phase, individual ${}^{85}\mathrm{Rb}$ atoms are probabilistically loaded into the workspace grid utilizing collisional blockade \cite{SchlosserN2001a,Grunzweig2010,Schlosser2012,Lester2015}.
\begin{figure*}[t]
	\includegraphics[width=\linewidth]{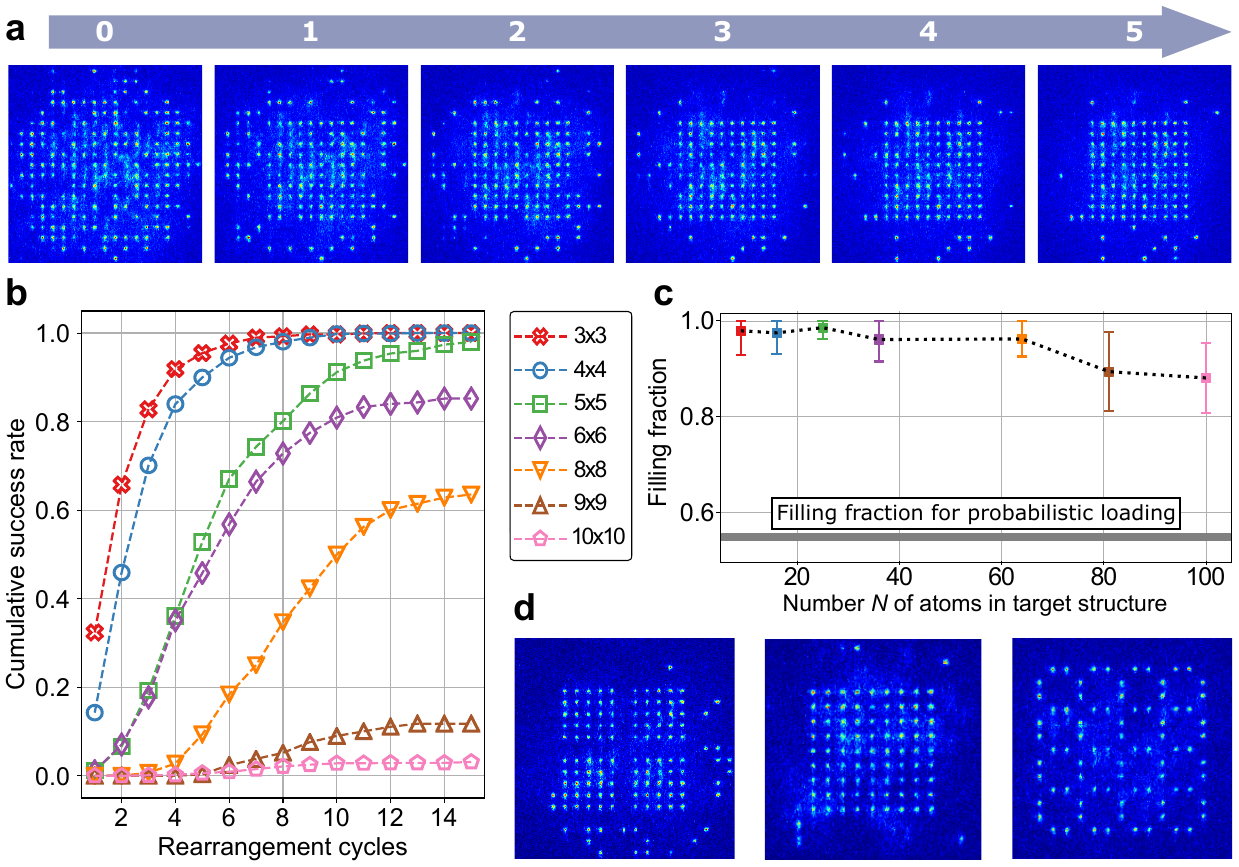}
	\caption{Multiple rearrangements leading to large clusters with high success rates and filling fractions. (a)  Atom distribution during a sequence of rearrangement cycles for a $10 \times 10$ target structure. Starting from an unsorted atom array, a defect-free cluster is generated within 5 cycles.
		(b) Measured cumulative success rates of achieving defect-free quadratic clusters of different sizes. For most clusters, the final value is reached after 10-15 rearrangement cycles.
		(c) Maximum filling fraction observed during rearrangement runs for different cluster sizes, which is typically reached after 7-9 cycles. Error bars correspond to a 1$\sigma$ interval. The thick continuous line at $0.55$ represents an upper bound to the filling fraction obtained via collisional blockade alone, i.e., the situation before the first rearrangement cycle.
		(d) Gallery of defect-free clusters suitable for quantum error correction and topological quantum computing. (Left) Four separate 25-atom clusters representing four logical surface-code qubits for single quantum-error correction \cite{Fowler2012}, (Middle) $9 \times 9$ atom cluster corresponding to one logical qubit for double-error correction, and (Right) 96-atom ring network as building block for an implementation of color-code schemes \cite{Bombin2008}.}
	\label{fig:cum_prob}
\end{figure*}
We determine the occupancy of the traps by fluorescence imaging with an integration time of $\unit{50-75}{\milli\second}$ and observe an average number of $191(17)$ trapped individual atoms. For each rearrangement cycle, we apply a shortest-move heuristic algorithm to calculate a sequence of atom moves to fill all vacant spots in a predefined target structure.
Atoms are moved along the virtual grid lines connecting the sites. If a calculated path contains an occupied trap along the way, the obstacle atom in that trap is moved into the target trap instead, with the original reservoir atom taking its place.
The algorithm attempts to optimize the transfer sequence by choosing the paths with the fewest obstacle atoms. Note that this algorithm does not necessarily find an optimal, yet time-efficient solution with future prospects to reduce the number of moves using trajectories bypassing occupied sites. During a rearrangement cycle, the calculated sequence of elementary rearrangement operations is carried out by the tweezer after the array depth is typically lowered by a factor of 4. A typical duration for a single transport is $\unit{1}{\milli\second}$. After the sequence, the regular array depth is reestablished and the resulting atomic positions are detected through fluorescence imaging.
We measure an average probability for successfully transporting a single atom of \unit{75}{\%} mainly limited by imperfect reloading of the atom from the tweezer to the target trap. In case of defects due to imperfect transport or lifetime-related atom losses, another rearrangement cycle attempts to eliminate them.
Furthermore, we observe a stabilizing effect for the atom lifetime due to laser cooling of the atoms during the repetitive fluorescence imaging, increasing the $1/e$ lifetime $\tau$ from a photon-scattering dominated value of $\unit{2.5}{\second}$ to a vacuum-limited value of $\unit{10}{\second}$.
Based on this procedure, \figref[c]{setup} shows the largest defect-free atom-by-atom-assembled structures reported so far containing up to 111 atoms in various configurations.\\
\figureref[]{cum_prob} documents the benefit of multiple rearrangement cycles \cite{Endres2016,Kim2017,Kumar2018} with respect to the scalability of the cluster size.
In \figref[a]{cum_prob}, a sequence for the assembly of a target structure of $10 \times 10$ atoms ($N=100$) is shown. The images depict the full series of consecutive stages of the assembly process, starting with the unsorted initial atom distribution. Between each image, a sequence of rearrangements is executed in order to reach defect-free filling, which is achieved after 5 cycles, taking $\unit{1.3}{\second}$ in total.
\\ \indent
In \figref[b]{cum_prob}, we show the cumulative success rates for defect-free assembly of quadratic target clusters of different sizes within a series of 15 rearrangement cycles. Up to a cluster size of $5 \times 5$ atoms, the cumulative success rate exceeds \unit{99}{\%}.
For larger clusters, atom losses during transport and lifetime-related losses out of the target structure prevent complete filling in every repetition of the experiment. Nevertheless, even the largest clusters have a high cumulative success rate, reaching values of \unit{64}{\%} for $8 \times 8$ atoms ($N = 64$), \unit{12}{\%} for $9 \times 9$ atoms ($N = 81$), and \unit{3.1}{\%} for $10 \times 10$ atoms ($N = 100$). When comparing the final values of the success rates to the ones after a single rearrangement cycle, the benefit of this method becomes evident: most of the curves saturate only after more than 10 rearrangement cycles. The maximum filling fraction of different quadratic clusters, as shown in \figref[c]{cum_prob} in analogy to \cite{Endres2016,Barredo2016,Kumar2018}, 
results in \unit{88(7)}{\%} for a $10 \times 10$ atom cluster and exceeds \unit{95}{\%} for all target clusters up to $8 \times 8$ atoms. Multitudinous rearrangement cycles are also essential for generating a specific set of atom clusters suitable for quantum error correction and topological quantum computing \cite{Bombin2008,Fowler2012}. \figureref[d]{cum_prob} presents a gallery of defect-free clusters that represent building blocks for these implementations (see caption of \figref[]{cum_prob} for details).
\begin{figure}[b]
	\includegraphics[width=\linewidth]{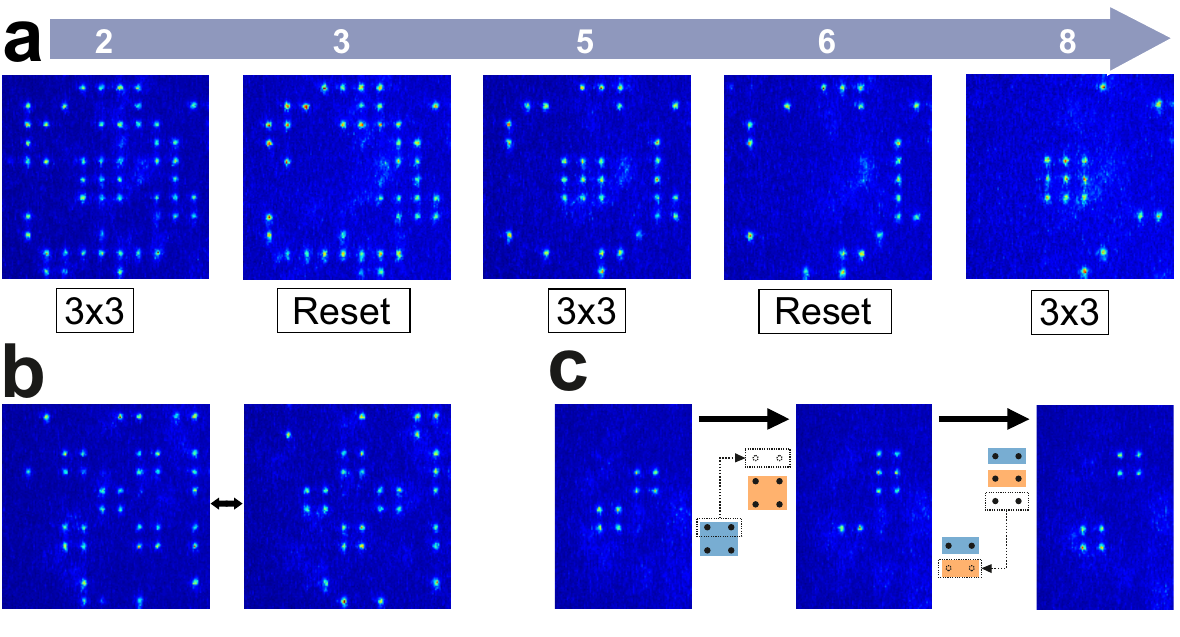}
	\caption{Demonstration of reloading and reordering schemes. (a) Multiple reconstruction of a central $3 \times 3$ cluster after deliberate atom removal. The number of feasible repetitions within one experimental run scales with the size of the reservoir array. This demonstration is based on an $11 \times 11$ workspace with 96 reservoir sites. 
		(b) Example of the transformation (inversion) of an atom arrangement within a single experimental run. Atoms lost during this procedure are replaced by atoms from the surrounding reservoir.
		(c) Demonstration of an atom exchange between two clusters: (Left) From a probabilistic initial atom distribution a defect-free structure of two $2 \times 2$ clusters is created. (Middle),(Right) The atoms are relocated so that two atoms of each cluster are moved into the respective other cluster. The two colors in the schematic correspond to the respective original clusters. This procedure will be used for the distribution of entanglement \cite{Kaufman2015}.}
	\label{fig:reloc}
\end{figure}
\\ \indent
Atom loss constitutes a major limitation in scalability, whether it is caused by experimental noise or intentional events, such as destructive state detection.
A reservoir of atoms outside the target structure can be used to heal emerging defects, mitigate the impact of losses, and significantly enhance the data acquisition rate by 
reducing the number of time-consuming atom loading and trapping phases. In \figref[a]{reloc}, we demonstrate the repeated reconstruction of a defect-free $3 \times 3$ cluster by intentionally emptying the target cluster and reloading atoms from a spatially separated reservoir.
Skipping the intentional removal of atoms, we have observed the perpetuation of a fully filled $5 \times 5$ cluster over the course of up to $\unit{10}{\second}$, by repeating a rearrangement cycle 80 times in a row. In 49(13) of the 80 images taken in this series, the target structure was determined to be without defect. With enough reservoir atoms, one can effectively extend the target cluster lifetime orders of magnitude beyond the one given by the atom loss rate.
Apart from reloading the same target structure multiple times, this technique also enables us to rearrange the atoms into different configurations within one experimental run, as is shown in \figref[b]{reloc}, where we switch between two inverted patterns.
Finally, deterministic atom transport allows for the transfer of particular atoms to new sites. Done adiabatically, the transfer preserves atomic coherence \cite{Lengwenus2010,Schlosser2012} and thus allows for the redistribution of quantum-correlated or entangled subarrays within large-scale atom clusters. As a proof of principle, in \figref[c]{reloc} we demonstrate a rearrangement sequence for four pairs of atoms  that could be used for the redistribution of entanglement \cite{Kaufman2015} between the two $2 \times 2$ atom clusters.
%
%
%
\\ \indent
In this Letter, we have presented a novel platform for the defect-free assembly of large-scale 2D atom clusters and demonstrated significant advances in success rates and maximum cluster size. Already in our current setup we achieve focal grids with up to several thousands of sites.
Only finite laser power and limited transport efficiency prevent us from working with arrays of several hundreds of traps and atoms.
A Monte Carlo simulation allows us to assess the full potential of our approach.
Based on realistically improved parameters, such as an increase in laser power to the maximum value commercially available, an initial loading rate of \unit{80}{\%} \cite{Brown2019,Grunzweig2010,Lester2015}, a vacuum-limited lifetime of $\tau = \unit{60}{\second}$ and a transport efficiency of \unit{95}{\%} (a value of \unit{99.3}{\%} has been reported in \cite{Barredo2016}), a simulation of rearrangements on a $50 \times 50$ grid yields a success rate $>\unit{90}{\%}$ for defect-free assembly of a 1000-atom target structure. 
With each site in the trap array corresponding to illuminating a separate lenslet in the MLA, in extended setups, microtrap arrays can be composed of multiple laser sources illuminating different sections of an extended MLA, further boosting scalability into the regime of $10^6$ trap sites.
Commercial MLAs with $1000 \times 1000$ microlenses have already been produced using lithographic manufacturing techniques.
\\ \indent
Our architecture lends itself to quantum metrology, simulation, and computation applications, including the implementation of topological quantum computing and quantum error correction \cite{Bombin2008,Fowler2012} based on Rydberg-mediated interactions \cite{SaffmanWalkerMolmer2010,Saffman2016,Browaeys2016,Levine2018}.
Reduced trap separations pave the way to bottom-up engineering of quantum systems based on tunneling interactions \cite{Kaufman2015,Murmann2015,Sturm2017}.
While all results presented here are based on a quadratic grid, hexagonal MLAs are readily available and direct laser writing methods give access to user-defined geometries \cite{Gissibl2016}.
Facilitated by the inherent self-imaging property of the 2D periodic optical trap array that creates a Talbot optical lattice \cite{Wen2013,Schlosser2019}, microlens generated single-atom arrays are expandable to three-dimensional multilayer configurations at no additional cost.

\begin{acknowledgments}
We acknowledge financial support from the Deutsche Forschungsgemeinschaft (DFG) [Grant No. BI 647/6-1, Priority Program SPP 1929 (GiRyd)]. We thank the \textsc{labscript suite} community for support in implementing state-of-the-art control software for our experiments.
D.O. and D.S. contributed equally to this work.
\end{acknowledgments}

\bibliographystyle{apsrev4-1}
\bibliography{100Atoms}

\begin{thebibliography}{38}%
\makeatletter
\providecommand \@ifxundefined [1]{%
 \@ifx{#1\undefined}
}%
\providecommand \@ifnum [1]{%
 \ifnum #1\expandafter \@firstoftwo
 \else \expandafter \@secondoftwo
 \fi
}%
\providecommand \@ifx [1]{%
 \ifx #1\expandafter \@firstoftwo
 \else \expandafter \@secondoftwo
 \fi
}%
\providecommand \natexlab [1]{#1}%
\providecommand \enquote  [1]{``#1''}%
\providecommand \bibnamefont  [1]{#1}%
\providecommand \bibfnamefont [1]{#1}%
\providecommand \citenamefont [1]{#1}%
\providecommand \href@noop [0]{\@secondoftwo}%
\providecommand \href [0]{\begingroup \@sanitize@url \@href}%
\providecommand \@href[1]{\@@startlink{#1}\@@href}%
\providecommand \@@href[1]{\endgroup#1\@@endlink}%
\providecommand \@sanitize@url [0]{\catcode `\\12\catcode `\$12\catcode
  `\&12\catcode `\#12\catcode `\^12\catcode `\_12\catcode `\%12\relax}%
\providecommand \@@startlink[1]{}%
\providecommand \@@endlink[0]{}%
\providecommand \url  [0]{\begingroup\@sanitize@url \@url }%
\providecommand \@url [1]{\endgroup\@href {#1}{\urlprefix }}%
\providecommand \urlprefix  [0]{URL }%
\providecommand \Eprint [0]{\href }%
\providecommand \doibase [0]{http://dx.doi.org/}%
\providecommand \selectlanguage [0]{\@gobble}%
\providecommand \bibinfo  [0]{\@secondoftwo}%
\providecommand \bibfield  [0]{\@secondoftwo}%
\providecommand \translation [1]{[#1]}%
\providecommand \BibitemOpen [0]{}%
\providecommand \bibitemStop [0]{}%
\providecommand \bibitemNoStop [0]{.\EOS\space}%
\providecommand \EOS [0]{\spacefactor3000\relax}%
\providecommand \BibitemShut  [1]{\csname bibitem#1\endcsname}%
\let\auto@bib@innerbib\@empty
\bibitem [{\citenamefont {Ac\'{i}n}\ \emph {et~al.}(2018)\citenamefont
  {Ac\'{i}n}, \citenamefont {Bloch}, \citenamefont {Buhrman}, \citenamefont
  {Calarco}, \citenamefont {Eichler}, \citenamefont {Eisert}, \citenamefont
  {Esteve}, \citenamefont {Gisin}, \citenamefont {Glaser}, \citenamefont
  {Jelezko}, \citenamefont {Kuhr}, \citenamefont {Lewenstein}, \citenamefont
  {Riedel}, \citenamefont {Schmidt}, \citenamefont {Thew}, \citenamefont
  {Wallraff}, \citenamefont {Walmsley},\ and\ \citenamefont
  {Wilhelm}}]{Acin2018}%
  \BibitemOpen
  \bibfield  {author} {\bibinfo {author} {\bibfnamefont {A.}~\bibnamefont
  {Ac\'{i}n}}, \bibinfo {author} {\bibfnamefont {I.}~\bibnamefont {Bloch}},
  \bibinfo {author} {\bibfnamefont {H.}~\bibnamefont {Buhrman}}, \bibinfo
  {author} {\bibfnamefont {T.}~\bibnamefont {Calarco}}, \bibinfo {author}
  {\bibfnamefont {C.}~\bibnamefont {Eichler}}, \bibinfo {author} {\bibfnamefont
  {J.}~\bibnamefont {Eisert}}, \bibinfo {author} {\bibfnamefont
  {D.}~\bibnamefont {Esteve}}, \bibinfo {author} {\bibfnamefont
  {N.}~\bibnamefont {Gisin}}, \bibinfo {author} {\bibfnamefont {S.~J.}\
  \bibnamefont {Glaser}}, \bibinfo {author} {\bibfnamefont {F.}~\bibnamefont
  {Jelezko}}, \bibinfo {author} {\bibfnamefont {S.}~\bibnamefont {Kuhr}},
  \bibinfo {author} {\bibfnamefont {M.}~\bibnamefont {Lewenstein}}, \bibinfo
  {author} {\bibfnamefont {M.~F.}\ \bibnamefont {Riedel}}, \bibinfo {author}
  {\bibfnamefont {P.~O.}\ \bibnamefont {Schmidt}}, \bibinfo {author}
  {\bibfnamefont {R.}~\bibnamefont {Thew}}, \bibinfo {author} {\bibfnamefont
  {A.}~\bibnamefont {Wallraff}}, \bibinfo {author} {\bibfnamefont
  {I.}~\bibnamefont {Walmsley}}, \ and\ \bibinfo {author} {\bibfnamefont
  {F.~K.}\ \bibnamefont {Wilhelm}},\ }\href
  {http://stacks.iop.org/1367-2630/20/i=8/a=080201} {\bibfield  {journal}
  {\bibinfo  {journal} {New J. Phys.}\ }\textbf {\bibinfo {volume} {20}},\
  \bibinfo {pages} {080201} (\bibinfo {year} {2018})}\BibitemShut {NoStop}%
\bibitem [{\citenamefont {Schlosser}\ \emph {et~al.}(2011)\citenamefont
  {Schlosser}, \citenamefont {Tichelmann}, \citenamefont {Kruse},\ and\
  \citenamefont {Birkl}}]{Schlosser2011}%
  \BibitemOpen
  \bibfield  {author} {\bibinfo {author} {\bibfnamefont {M.}~\bibnamefont
  {Schlosser}}, \bibinfo {author} {\bibfnamefont {S.}~\bibnamefont
  {Tichelmann}}, \bibinfo {author} {\bibfnamefont {J.}~\bibnamefont {Kruse}}, \
  and\ \bibinfo {author} {\bibfnamefont {G.}~\bibnamefont {Birkl}},\ }\href
  {\doibase 10.1007/s11128-011-0297-z} {\bibfield  {journal} {\bibinfo
  {journal} {Quantum Inf. Process.}\ }\textbf {\bibinfo {volume} {10}},\
  \bibinfo {pages} {907} (\bibinfo {year} {2011})}\BibitemShut {NoStop}%
\bibitem [{\citenamefont {Piotrowicz}\ \emph {et~al.}(2013)\citenamefont
  {Piotrowicz}, \citenamefont {Lichtman}, \citenamefont {Maller}, \citenamefont
  {Li}, \citenamefont {Zhang}, \citenamefont {Isenhower},\ and\ \citenamefont
  {Saffman}}]{Piotrowicz2013}%
  \BibitemOpen
  \bibfield  {author} {\bibinfo {author} {\bibfnamefont {M.~J.}\ \bibnamefont
  {Piotrowicz}}, \bibinfo {author} {\bibfnamefont {M.}~\bibnamefont
  {Lichtman}}, \bibinfo {author} {\bibfnamefont {K.}~\bibnamefont {Maller}},
  \bibinfo {author} {\bibfnamefont {G.}~\bibnamefont {Li}}, \bibinfo {author}
  {\bibfnamefont {S.}~\bibnamefont {Zhang}}, \bibinfo {author} {\bibfnamefont
  {L.}~\bibnamefont {Isenhower}}, \ and\ \bibinfo {author} {\bibfnamefont
  {M.}~\bibnamefont {Saffman}},\ }\href {\doibase 10.1103/PhysRevA.88.013420}
  {\bibfield  {journal} {\bibinfo  {journal} {Phys. Rev. A}\ }\textbf {\bibinfo
  {volume} {88}},\ \bibinfo {pages} {013420} (\bibinfo {year}
  {2013})}\BibitemShut {NoStop}%
\bibitem [{\citenamefont {Endres}\ \emph {et~al.}(2016)\citenamefont {Endres},
  \citenamefont {Bernien}, \citenamefont {Keesling}, \citenamefont {Levine},
  \citenamefont {Anschuetz}, \citenamefont {Krajenbrink}, \citenamefont
  {Senko}, \citenamefont {Vuletic}, \citenamefont {Greiner},\ and\
  \citenamefont {Lukin}}]{Endres2016}%
  \BibitemOpen
  \bibfield  {author} {\bibinfo {author} {\bibfnamefont {M.}~\bibnamefont
  {Endres}}, \bibinfo {author} {\bibfnamefont {H.}~\bibnamefont {Bernien}},
  \bibinfo {author} {\bibfnamefont {A.}~\bibnamefont {Keesling}}, \bibinfo
  {author} {\bibfnamefont {H.}~\bibnamefont {Levine}}, \bibinfo {author}
  {\bibfnamefont {E.~R.}\ \bibnamefont {Anschuetz}}, \bibinfo {author}
  {\bibfnamefont {A.}~\bibnamefont {Krajenbrink}}, \bibinfo {author}
  {\bibfnamefont {C.}~\bibnamefont {Senko}}, \bibinfo {author} {\bibfnamefont
  {V.}~\bibnamefont {Vuletic}}, \bibinfo {author} {\bibfnamefont
  {M.}~\bibnamefont {Greiner}}, \ and\ \bibinfo {author} {\bibfnamefont
  {M.~D.}\ \bibnamefont {Lukin}},\ }\href {\doibase 10.1126/science.aah3752}
  {\bibfield  {journal} {\bibinfo  {journal} {Science}\ }\textbf {\bibinfo
  {volume} {354}},\ \bibinfo {pages} {1024} (\bibinfo {year}
  {2016})}\BibitemShut {NoStop}%
\bibitem [{\citenamefont {Barredo}\ \emph {et~al.}(2016)\citenamefont
  {Barredo}, \citenamefont {de~L{\'e}s{\'e}leuc}, \citenamefont {Lienhard},
  \citenamefont {Lahaye},\ and\ \citenamefont {Browaeys}}]{Barredo2016}%
  \BibitemOpen
  \bibfield  {author} {\bibinfo {author} {\bibfnamefont {D.}~\bibnamefont
  {Barredo}}, \bibinfo {author} {\bibfnamefont {S.}~\bibnamefont
  {de~L{\'e}s{\'e}leuc}}, \bibinfo {author} {\bibfnamefont {V.}~\bibnamefont
  {Lienhard}}, \bibinfo {author} {\bibfnamefont {T.}~\bibnamefont {Lahaye}}, \
  and\ \bibinfo {author} {\bibfnamefont {A.}~\bibnamefont {Browaeys}},\ }\href
  {\doibase 10.1126/science.aah3778} {\bibfield  {journal} {\bibinfo  {journal}
  {Science}\ }\textbf {\bibinfo {volume} {354}},\ \bibinfo {pages} {1021}
  (\bibinfo {year} {2016})}\BibitemShut {NoStop}%
\bibitem [{\citenamefont {Kim}\ \emph {et~al.}(2016)\citenamefont {Kim},
  \citenamefont {Lee}, \citenamefont {Lee}, \citenamefont {Jo}, \citenamefont
  {Song},\ and\ \citenamefont {Ahn}}]{Kim2016}%
  \BibitemOpen
  \bibfield  {author} {\bibinfo {author} {\bibfnamefont {H.}~\bibnamefont
  {Kim}}, \bibinfo {author} {\bibfnamefont {W.}~\bibnamefont {Lee}}, \bibinfo
  {author} {\bibfnamefont {H.-g.}\ \bibnamefont {Lee}}, \bibinfo {author}
  {\bibfnamefont {H.}~\bibnamefont {Jo}}, \bibinfo {author} {\bibfnamefont
  {Y.}~\bibnamefont {Song}}, \ and\ \bibinfo {author} {\bibfnamefont
  {J.}~\bibnamefont {Ahn}},\ }\href {http://dx.doi.org/10.1038/ncomms13317}
  {\bibfield  {journal} {\bibinfo  {journal} {Nat. Commun.}\ }\textbf {\bibinfo
  {volume} {7}},\ \bibinfo {pages} {13317} (\bibinfo {year}
  {2016})}\BibitemShut {NoStop}%
\bibitem [{\citenamefont {Gross}\ and\ \citenamefont
  {Bloch}(2017)}]{Gross2017}%
  \BibitemOpen
  \bibfield  {author} {\bibinfo {author} {\bibfnamefont {C.}~\bibnamefont
  {Gross}}\ and\ \bibinfo {author} {\bibfnamefont {I.}~\bibnamefont {Bloch}},\
  }\href {\doibase 10.1126/science.aal3837} {\bibfield  {journal} {\bibinfo
  {journal} {Science}\ }\textbf {\bibinfo {volume} {357}},\ \bibinfo {pages}
  {995} (\bibinfo {year} {2017})}\BibitemShut {NoStop}%
\bibitem [{\citenamefont {Barredo}\ \emph {et~al.}(2018)\citenamefont
  {Barredo}, \citenamefont {Lienhard}, \citenamefont {De~L\'{e}s\'{e}leuc},
  \citenamefont {Lahaye},\ and\ \citenamefont {Browaeys}}]{Barredo2018}%
  \BibitemOpen
  \bibfield  {author} {\bibinfo {author} {\bibfnamefont {D.}~\bibnamefont
  {Barredo}}, \bibinfo {author} {\bibfnamefont {V.}~\bibnamefont {Lienhard}},
  \bibinfo {author} {\bibfnamefont {S.}~\bibnamefont {De~L\'{e}s\'{e}leuc}},
  \bibinfo {author} {\bibfnamefont {T.}~\bibnamefont {Lahaye}}, \ and\ \bibinfo
  {author} {\bibfnamefont {A.}~\bibnamefont {Browaeys}},\ }\href {\doibase
  10.1038/s41586-018-0450-2} {\bibfield  {journal} {\bibinfo  {journal}
  {Nature}\ }\textbf {\bibinfo {volume} {561}},\ \bibinfo {pages} {79}
  (\bibinfo {year} {2018})}\BibitemShut {NoStop}%
\bibitem [{\citenamefont {Kumar}\ \emph {et~al.}(2018)\citenamefont {Kumar},
  \citenamefont {Wu}, \citenamefont {Giraldo},\ and\ \citenamefont
  {Weiss}}]{Kumar2018}%
  \BibitemOpen
  \bibfield  {author} {\bibinfo {author} {\bibfnamefont {A.}~\bibnamefont
  {Kumar}}, \bibinfo {author} {\bibfnamefont {T.-Y.}\ \bibnamefont {Wu}},
  \bibinfo {author} {\bibfnamefont {F.}~\bibnamefont {Giraldo}}, \ and\
  \bibinfo {author} {\bibfnamefont {D.~S.}\ \bibnamefont {Weiss}},\ }\href
  {\doibase 10.1038/s41586-018-0458-7} {\bibfield  {journal} {\bibinfo
  {journal} {Nature}\ }\textbf {\bibinfo {volume} {561}},\ \bibinfo {pages}
  {83} (\bibinfo {year} {2018})}\BibitemShut {NoStop}%
\bibitem [{\citenamefont {Brown}\ \emph {et~al.}(2019)\citenamefont {Brown},
  \citenamefont {Thiele}, \citenamefont {Kiehl}, \citenamefont {Hsu},\ and\
  \citenamefont {Regal}}]{Brown2019}%
  \BibitemOpen
  \bibfield  {author} {\bibinfo {author} {\bibfnamefont {M.~O.}\ \bibnamefont
  {Brown}}, \bibinfo {author} {\bibfnamefont {T.}~\bibnamefont {Thiele}},
  \bibinfo {author} {\bibfnamefont {C.}~\bibnamefont {Kiehl}}, \bibinfo
  {author} {\bibfnamefont {T.-W.}\ \bibnamefont {Hsu}}, \ and\ \bibinfo
  {author} {\bibfnamefont {C.~A.}\ \bibnamefont {Regal}},\ }\href {\doibase
  10.1103/PhysRevX.9.011057} {\bibfield  {journal} {\bibinfo  {journal} {Phys.
  Rev. X}\ }\textbf {\bibinfo {volume} {9}},\ \bibinfo {pages} {011057}
  (\bibinfo {year} {2019})}\BibitemShut {NoStop}%
\bibitem [{\citenamefont {Saffman}\ \emph {et~al.}(2010)\citenamefont
  {Saffman}, \citenamefont {Walker},\ and\ \citenamefont
  {M\o{}lmer}}]{SaffmanWalkerMolmer2010}%
  \BibitemOpen
  \bibfield  {author} {\bibinfo {author} {\bibfnamefont {M.}~\bibnamefont
  {Saffman}}, \bibinfo {author} {\bibfnamefont {T.~G.}\ \bibnamefont {Walker}},
  \ and\ \bibinfo {author} {\bibfnamefont {K.}~\bibnamefont {M\o{}lmer}},\
  }\href {\doibase 10.1103/RevModPhys.82.2313} {\bibfield  {journal} {\bibinfo
  {journal} {Rev. Mod. Phys.}\ }\textbf {\bibinfo {volume} {82}},\ \bibinfo
  {pages} {2313} (\bibinfo {year} {2010})}\BibitemShut {NoStop}%
\bibitem [{\citenamefont {Saffman}(2016)}]{Saffman2016}%
  \BibitemOpen
  \bibfield  {author} {\bibinfo {author} {\bibfnamefont {M.}~\bibnamefont
  {Saffman}},\ }\href {\doibase 10.1088/0953-4075/49/20/202001} {\bibfield
  {journal} {\bibinfo  {journal} {J. Phys. B}\ }\textbf {\bibinfo {volume}
  {49}},\ \bibinfo {pages} {202001} (\bibinfo {year} {2016})}\BibitemShut
  {NoStop}%
\bibitem [{\citenamefont {Browaeys}\ \emph {et~al.}(2016)\citenamefont
  {Browaeys}, \citenamefont {Barredo},\ and\ \citenamefont
  {Lahaye}}]{Browaeys2016}%
  \BibitemOpen
  \bibfield  {author} {\bibinfo {author} {\bibfnamefont {A.}~\bibnamefont
  {Browaeys}}, \bibinfo {author} {\bibfnamefont {D.}~\bibnamefont {Barredo}}, \
  and\ \bibinfo {author} {\bibfnamefont {T.}~\bibnamefont {Lahaye}},\ }\href
  {http://stacks.iop.org/0953-4075/49/i=15/a=152001} {\bibfield  {journal}
  {\bibinfo  {journal} {J. Phys. B}\ }\textbf {\bibinfo {volume} {49}},\
  \bibinfo {pages} {152001} (\bibinfo {year} {2016})}\BibitemShut {NoStop}%
\bibitem [{\citenamefont {Bakr}\ \emph {et~al.}(2010)\citenamefont {Bakr},
  \citenamefont {Peng}, \citenamefont {Tai}, \citenamefont {Ma}, \citenamefont
  {Simon}, \citenamefont {Gillen}, \citenamefont {F{\"o}lling}, \citenamefont
  {Pollet},\ and\ \citenamefont {Greiner}}]{Bakr2010}%
  \BibitemOpen
  \bibfield  {author} {\bibinfo {author} {\bibfnamefont {W.~S.}\ \bibnamefont
  {Bakr}}, \bibinfo {author} {\bibfnamefont {A.}~\bibnamefont {Peng}}, \bibinfo
  {author} {\bibfnamefont {M.~E.}\ \bibnamefont {Tai}}, \bibinfo {author}
  {\bibfnamefont {R.}~\bibnamefont {Ma}}, \bibinfo {author} {\bibfnamefont
  {J.}~\bibnamefont {Simon}}, \bibinfo {author} {\bibfnamefont {J.~I.}\
  \bibnamefont {Gillen}}, \bibinfo {author} {\bibfnamefont {S.}~\bibnamefont
  {F{\"o}lling}}, \bibinfo {author} {\bibfnamefont {L.}~\bibnamefont {Pollet}},
  \ and\ \bibinfo {author} {\bibfnamefont {M.}~\bibnamefont {Greiner}},\ }\href
  {\doibase 10.1126/science.1192368} {\bibfield  {journal} {\bibinfo  {journal}
  {Science}\ }\textbf {\bibinfo {volume} {329}},\ \bibinfo {pages} {547}
  (\bibinfo {year} {2010})}\BibitemShut {NoStop}%
\bibitem [{\citenamefont {Sherson}\ \emph {et~al.}(2010)\citenamefont
  {Sherson}, \citenamefont {Weitenberg}, \citenamefont {Endres}, \citenamefont
  {Cheneau}, \citenamefont {Bloch},\ and\ \citenamefont {Kuhr}}]{Sherson2010}%
  \BibitemOpen
  \bibfield  {author} {\bibinfo {author} {\bibfnamefont {J.~F.}\ \bibnamefont
  {Sherson}}, \bibinfo {author} {\bibfnamefont {C.}~\bibnamefont {Weitenberg}},
  \bibinfo {author} {\bibfnamefont {M.}~\bibnamefont {Endres}}, \bibinfo
  {author} {\bibfnamefont {M.}~\bibnamefont {Cheneau}}, \bibinfo {author}
  {\bibfnamefont {I.}~\bibnamefont {Bloch}}, \ and\ \bibinfo {author}
  {\bibfnamefont {S.}~\bibnamefont {Kuhr}},\ }\href {\doibase
  10.1038/nature09378} {\bibfield  {journal} {\bibinfo  {journal} {Nature}\
  }\textbf {\bibinfo {volume} {467}},\ \bibinfo {pages} {68} (\bibinfo {year}
  {2010})}\BibitemShut {NoStop}%
\bibitem [{\citenamefont {Zeiher}\ \emph {et~al.}(2015)\citenamefont {Zeiher},
  \citenamefont {Schau\ss{}}, \citenamefont {Hild}, \citenamefont {Macr\`{\i}},
  \citenamefont {Bloch},\ and\ \citenamefont {Gross}}]{Zeiher2015}%
  \BibitemOpen
  \bibfield  {author} {\bibinfo {author} {\bibfnamefont {J.}~\bibnamefont
  {Zeiher}}, \bibinfo {author} {\bibfnamefont {P.}~\bibnamefont {Schau\ss{}}},
  \bibinfo {author} {\bibfnamefont {S.}~\bibnamefont {Hild}}, \bibinfo {author}
  {\bibfnamefont {T.}~\bibnamefont {Macr\`{\i}}}, \bibinfo {author}
  {\bibfnamefont {I.}~\bibnamefont {Bloch}}, \ and\ \bibinfo {author}
  {\bibfnamefont {C.}~\bibnamefont {Gross}},\ }\href {\doibase
  10.1103/PhysRevX.5.031015} {\bibfield  {journal} {\bibinfo  {journal} {Phys.
  Rev. X}\ }\textbf {\bibinfo {volume} {5}},\ \bibinfo {pages} {031015}
  (\bibinfo {year} {2015})}\BibitemShut {NoStop}%
\bibitem [{\citenamefont {Mazurenko}\ \emph {et~al.}(2017)\citenamefont
  {Mazurenko}, \citenamefont {Chiu}, \citenamefont {Ji}, \citenamefont
  {Parsons}, \citenamefont {Kan{\'a}sz-Nagy}, \citenamefont {Schmidt},
  \citenamefont {Grusdt}, \citenamefont {Demler}, \citenamefont {Greif},\ and\
  \citenamefont {Greiner}}]{Mazurenko2017}%
  \BibitemOpen
  \bibfield  {author} {\bibinfo {author} {\bibfnamefont {A.}~\bibnamefont
  {Mazurenko}}, \bibinfo {author} {\bibfnamefont {C.~S.}\ \bibnamefont {Chiu}},
  \bibinfo {author} {\bibfnamefont {G.}~\bibnamefont {Ji}}, \bibinfo {author}
  {\bibfnamefont {M.~F.}\ \bibnamefont {Parsons}}, \bibinfo {author}
  {\bibfnamefont {M.}~\bibnamefont {Kan{\'a}sz-Nagy}}, \bibinfo {author}
  {\bibfnamefont {R.}~\bibnamefont {Schmidt}}, \bibinfo {author} {\bibfnamefont
  {F.}~\bibnamefont {Grusdt}}, \bibinfo {author} {\bibfnamefont
  {E.}~\bibnamefont {Demler}}, \bibinfo {author} {\bibfnamefont
  {D.}~\bibnamefont {Greif}}, \ and\ \bibinfo {author} {\bibfnamefont
  {M.}~\bibnamefont {Greiner}},\ }\href {\doibase 10.1038/nature22362}
  {\bibfield  {journal} {\bibinfo  {journal} {Nature}\ }\textbf {\bibinfo
  {volume} {545}},\ \bibinfo {pages} {462} (\bibinfo {year}
  {2017})}\BibitemShut {NoStop}%
\bibitem [{\citenamefont {Chiu}\ \emph {et~al.}(2018)\citenamefont {Chiu},
  \citenamefont {Ji}, \citenamefont {Mazurenko}, \citenamefont {Greif},\ and\
  \citenamefont {Greiner}}]{Chiu2018}%
  \BibitemOpen
  \bibfield  {author} {\bibinfo {author} {\bibfnamefont {C.~S.}\ \bibnamefont
  {Chiu}}, \bibinfo {author} {\bibfnamefont {G.}~\bibnamefont {Ji}}, \bibinfo
  {author} {\bibfnamefont {A.}~\bibnamefont {Mazurenko}}, \bibinfo {author}
  {\bibfnamefont {D.}~\bibnamefont {Greif}}, \ and\ \bibinfo {author}
  {\bibfnamefont {M.}~\bibnamefont {Greiner}},\ }\href {\doibase
  10.1103/PhysRevLett.120.243201} {\bibfield  {journal} {\bibinfo  {journal}
  {Phys. Rev. Lett.}\ }\textbf {\bibinfo {volume} {120}},\ \bibinfo {pages}
  {243201} (\bibinfo {year} {2018})}\BibitemShut {NoStop}%
\bibitem [{\citenamefont {Guardado-Sanchez}\ \emph {et~al.}(2018)\citenamefont
  {Guardado-Sanchez}, \citenamefont {Brown}, \citenamefont {Mitra},
  \citenamefont {Devakul}, \citenamefont {Huse}, \citenamefont {Schau\ss{}},\
  and\ \citenamefont {Bakr}}]{Guardado2018}%
  \BibitemOpen
  \bibfield  {author} {\bibinfo {author} {\bibfnamefont {E.}~\bibnamefont
  {Guardado-Sanchez}}, \bibinfo {author} {\bibfnamefont {P.~T.}\ \bibnamefont
  {Brown}}, \bibinfo {author} {\bibfnamefont {D.}~\bibnamefont {Mitra}},
  \bibinfo {author} {\bibfnamefont {T.}~\bibnamefont {Devakul}}, \bibinfo
  {author} {\bibfnamefont {D.~A.}\ \bibnamefont {Huse}}, \bibinfo {author}
  {\bibfnamefont {P.}~\bibnamefont {Schau\ss{}}}, \ and\ \bibinfo {author}
  {\bibfnamefont {W.~S.}\ \bibnamefont {Bakr}},\ }\href {\doibase
  10.1103/PhysRevX.8.021069} {\bibfield  {journal} {\bibinfo  {journal} {Phys.
  Rev. X}\ }\textbf {\bibinfo {volume} {8}},\ \bibinfo {pages} {021069}
  (\bibinfo {year} {2018})}\BibitemShut {NoStop}%
\bibitem [{\citenamefont {Robens}\ \emph {et~al.}(2017)\citenamefont {Robens},
  \citenamefont {Zopes}, \citenamefont {Alt}, \citenamefont {Brakhane},
  \citenamefont {Meschede},\ and\ \citenamefont {Alberti}}]{Robens2017}%
  \BibitemOpen
  \bibfield  {author} {\bibinfo {author} {\bibfnamefont {C.}~\bibnamefont
  {Robens}}, \bibinfo {author} {\bibfnamefont {J.}~\bibnamefont {Zopes}},
  \bibinfo {author} {\bibfnamefont {W.}~\bibnamefont {Alt}}, \bibinfo {author}
  {\bibfnamefont {S.}~\bibnamefont {Brakhane}}, \bibinfo {author}
  {\bibfnamefont {D.}~\bibnamefont {Meschede}}, \ and\ \bibinfo {author}
  {\bibfnamefont {A.}~\bibnamefont {Alberti}},\ }\href {\doibase
  10.1103/PhysRevLett.118.065302} {\bibfield  {journal} {\bibinfo  {journal}
  {Phys. Rev. Lett.}\ }\textbf {\bibinfo {volume} {118}},\ \bibinfo {pages}
  {065302} (\bibinfo {year} {2017})}\BibitemShut {NoStop}%
\bibitem [{\citenamefont {{Schlosser}}\ \emph {et~al.}(2001)\citenamefont
  {{Schlosser}}, \citenamefont {{Reymond}}, \citenamefont {{Protsenko}},\ and\
  \citenamefont {{Grangier}}}]{SchlosserN2001a}%
  \BibitemOpen
  \bibfield  {author} {\bibinfo {author} {\bibfnamefont {N.}~\bibnamefont
  {{Schlosser}}}, \bibinfo {author} {\bibfnamefont {G.}~\bibnamefont
  {{Reymond}}}, \bibinfo {author} {\bibfnamefont {I.}~\bibnamefont
  {{Protsenko}}}, \ and\ \bibinfo {author} {\bibfnamefont {P.}~\bibnamefont
  {{Grangier}}},\ }\href {\doibase 10.1038/35082512} {\bibfield  {journal}
  {\bibinfo  {journal} {Nature}\ }\textbf {\bibinfo {volume} {411}},\ \bibinfo
  {pages} {1024} (\bibinfo {year} {2001})}\BibitemShut {NoStop}%
\bibitem [{\citenamefont {{Gr{\"u}nzweig}}\ \emph {et~al.}(2010)\citenamefont
  {{Gr{\"u}nzweig}}, \citenamefont {{Hilliard}}, \citenamefont {{McGovern}},\
  and\ \citenamefont {{Andersen}}}]{Grunzweig2010}%
  \BibitemOpen
  \bibfield  {author} {\bibinfo {author} {\bibfnamefont {T.}~\bibnamefont
  {{Gr{\"u}nzweig}}}, \bibinfo {author} {\bibfnamefont {A.}~\bibnamefont
  {{Hilliard}}}, \bibinfo {author} {\bibfnamefont {M.}~\bibnamefont
  {{McGovern}}}, \ and\ \bibinfo {author} {\bibfnamefont {M.~F.}\ \bibnamefont
  {{Andersen}}},\ }\href {\doibase 10.1038/nphys1778} {\bibfield  {journal}
  {\bibinfo  {journal} {Nat. Phys.}\ }\textbf {\bibinfo {volume} {6}},\
  \bibinfo {pages} {951} (\bibinfo {year} {2010})}\BibitemShut {NoStop}%
\bibitem [{\citenamefont {Schlosser}\ \emph {et~al.}(2012)\citenamefont
  {Schlosser}, \citenamefont {Kruse}, \citenamefont {Gierl}, \citenamefont
  {Teichmann}, \citenamefont {Tichelmann},\ and\ \citenamefont
  {Birkl}}]{Schlosser2012}%
  \BibitemOpen
  \bibfield  {author} {\bibinfo {author} {\bibfnamefont {M.}~\bibnamefont
  {Schlosser}}, \bibinfo {author} {\bibfnamefont {J.}~\bibnamefont {Kruse}},
  \bibinfo {author} {\bibfnamefont {C.}~\bibnamefont {Gierl}}, \bibinfo
  {author} {\bibfnamefont {S.}~\bibnamefont {Teichmann}}, \bibinfo {author}
  {\bibfnamefont {S.}~\bibnamefont {Tichelmann}}, \ and\ \bibinfo {author}
  {\bibfnamefont {G.}~\bibnamefont {Birkl}},\ }\href {\doibase
  10.1088/1367-2630/14/12/123034} {\bibfield  {journal} {\bibinfo  {journal}
  {New J. Phys.}\ }\textbf {\bibinfo {volume} {14}},\ \bibinfo {pages} {123034}
  (\bibinfo {year} {2012})}\BibitemShut {NoStop}%
\bibitem [{\citenamefont {Lester}\ \emph {et~al.}(2015)\citenamefont {Lester},
  \citenamefont {Luick}, \citenamefont {Kaufman}, \citenamefont {Reynolds},\
  and\ \citenamefont {Regal}}]{Lester2015}%
  \BibitemOpen
  \bibfield  {author} {\bibinfo {author} {\bibfnamefont {B.~J.}\ \bibnamefont
  {Lester}}, \bibinfo {author} {\bibfnamefont {N.}~\bibnamefont {Luick}},
  \bibinfo {author} {\bibfnamefont {A.~M.}\ \bibnamefont {Kaufman}}, \bibinfo
  {author} {\bibfnamefont {C.~M.}\ \bibnamefont {Reynolds}}, \ and\ \bibinfo
  {author} {\bibfnamefont {C.~A.}\ \bibnamefont {Regal}},\ }\href {\doibase
  10.1103/PhysRevLett.115.073003} {\bibfield  {journal} {\bibinfo  {journal}
  {Phys. Rev. Lett.}\ }\textbf {\bibinfo {volume} {115}},\ \bibinfo {pages}
  {073003} (\bibinfo {year} {2015})}\BibitemShut {NoStop}%
\bibitem [{\citenamefont {Bernien}\ \emph {et~al.}(2017)\citenamefont
  {Bernien}, \citenamefont {Schwartz}, \citenamefont {Keesling}, \citenamefont
  {Levine}, \citenamefont {Omran}, \citenamefont {Pichler}, \citenamefont
  {Choi}, \citenamefont {Zibrov}, \citenamefont {Endres}, \citenamefont
  {Greiner} \emph {et~al.}}]{Bernien2017}%
  \BibitemOpen
  \bibfield  {author} {\bibinfo {author} {\bibfnamefont {H.}~\bibnamefont
  {Bernien}}, \bibinfo {author} {\bibfnamefont {S.}~\bibnamefont {Schwartz}},
  \bibinfo {author} {\bibfnamefont {A.}~\bibnamefont {Keesling}}, \bibinfo
  {author} {\bibfnamefont {H.}~\bibnamefont {Levine}}, \bibinfo {author}
  {\bibfnamefont {A.}~\bibnamefont {Omran}}, \bibinfo {author} {\bibfnamefont
  {H.}~\bibnamefont {Pichler}}, \bibinfo {author} {\bibfnamefont
  {S.}~\bibnamefont {Choi}}, \bibinfo {author} {\bibfnamefont {A.~S.}\
  \bibnamefont {Zibrov}}, \bibinfo {author} {\bibfnamefont {M.}~\bibnamefont
  {Endres}}, \bibinfo {author} {\bibfnamefont {M.}~\bibnamefont {Greiner}},
  \emph {et~al.},\ }\href {\doibase 10.1038/nature24622} {\bibfield  {journal}
  {\bibinfo  {journal} {Nature}\ }\textbf {\bibinfo {volume} {551}},\ \bibinfo
  {pages} {579} (\bibinfo {year} {2017})}\BibitemShut {NoStop}%
\bibitem [{\citenamefont {Kim}\ \emph {et~al.}(2018)\citenamefont {Kim},
  \citenamefont {Park}, \citenamefont {Kim}, \citenamefont {Sim},\ and\
  \citenamefont {Ahn}}]{Kim2018}%
  \BibitemOpen
  \bibfield  {author} {\bibinfo {author} {\bibfnamefont {H.}~\bibnamefont
  {Kim}}, \bibinfo {author} {\bibfnamefont {Y.~J.}\ \bibnamefont {Park}},
  \bibinfo {author} {\bibfnamefont {K.}~\bibnamefont {Kim}}, \bibinfo {author}
  {\bibfnamefont {H.-S.}\ \bibnamefont {Sim}}, \ and\ \bibinfo {author}
  {\bibfnamefont {J.}~\bibnamefont {Ahn}},\ }\href {\doibase
  10.1103/PhysRevLett.120.180502} {\bibfield  {journal} {\bibinfo  {journal}
  {Phys. Rev. Lett.}\ }\textbf {\bibinfo {volume} {120}},\ \bibinfo {pages}
  {180502} (\bibinfo {year} {2018})}\BibitemShut {NoStop}%
\bibitem [{\citenamefont {de~L{\'e}s{\'e}leuc}\ \emph {et~al.}()\citenamefont
  {de~L{\'e}s{\'e}leuc}, \citenamefont {Lienhard}, \citenamefont {Scholl},
  \citenamefont {Barredo}, \citenamefont {Weber}, \citenamefont {Lang},
  \citenamefont {B{\"u}chler}, \citenamefont {Lahaye},\ and\ \citenamefont
  {Browaeys}}]{Leseleuc2018}%
  \BibitemOpen
  \bibfield  {author} {\bibinfo {author} {\bibfnamefont {S.}~\bibnamefont
  {de~L{\'e}s{\'e}leuc}}, \bibinfo {author} {\bibfnamefont {V.}~\bibnamefont
  {Lienhard}}, \bibinfo {author} {\bibfnamefont {P.}~\bibnamefont {Scholl}},
  \bibinfo {author} {\bibfnamefont {D.}~\bibnamefont {Barredo}}, \bibinfo
  {author} {\bibfnamefont {S.}~\bibnamefont {Weber}}, \bibinfo {author}
  {\bibfnamefont {N.}~\bibnamefont {Lang}}, \bibinfo {author} {\bibfnamefont
  {H.~P.}\ \bibnamefont {B{\"u}chler}}, \bibinfo {author} {\bibfnamefont
  {T.}~\bibnamefont {Lahaye}}, \ and\ \bibinfo {author} {\bibfnamefont
  {A.}~\bibnamefont {Browaeys}},\ }\href {http://arxiv.org/abs/1810.13286}
  {\bibinfo  {journal} {arXiv:1810.13286}\ }\BibitemShut {NoStop}%
\bibitem [{\citenamefont {Fowler}\ \emph {et~al.}(2012)\citenamefont {Fowler},
  \citenamefont {Mariantoni}, \citenamefont {Martinis},\ and\ \citenamefont
  {Cleland}}]{Fowler2012}%
  \BibitemOpen
\bibfield  {journal} {  }\bibfield  {author} {\bibinfo {author} {\bibfnamefont
  {A.~G.}\ \bibnamefont {Fowler}}, \bibinfo {author} {\bibfnamefont
  {M.}~\bibnamefont {Mariantoni}}, \bibinfo {author} {\bibfnamefont {J.~M.}\
  \bibnamefont {Martinis}}, \ and\ \bibinfo {author} {\bibfnamefont {A.~N.}\
  \bibnamefont {Cleland}},\ }\href {\doibase 10.1103/PhysRevA.86.032324}
  {\bibfield  {journal} {\bibinfo  {journal} {Phys. Rev. A}\ }\textbf {\bibinfo
  {volume} {86}},\ \bibinfo {pages} {032324} (\bibinfo {year}
  {2012})}\BibitemShut {NoStop}%
\bibitem [{\citenamefont {Bombin}\ and\ \citenamefont
  {Martin-Delgado}(2008)}]{Bombin2008}%
  \BibitemOpen
  \bibfield  {author} {\bibinfo {author} {\bibfnamefont {H.}~\bibnamefont
  {Bombin}}\ and\ \bibinfo {author} {\bibfnamefont {M.~A.}\ \bibnamefont
  {Martin-Delgado}},\ }\href {\doibase 10.1103/PhysRevA.77.042322} {\bibfield
  {journal} {\bibinfo  {journal} {Phys. Rev. A}\ }\textbf {\bibinfo {volume}
  {77}},\ \bibinfo {pages} {042322} (\bibinfo {year} {2008})}\BibitemShut
  {NoStop}%
\bibitem [{\citenamefont {Lee}\ \emph {et~al.}(2017)\citenamefont {Lee},
  \citenamefont {Kim},\ and\ \citenamefont {Ahn}}]{Kim2017}%
  \BibitemOpen
  \bibfield  {author} {\bibinfo {author} {\bibfnamefont {W.}~\bibnamefont
  {Lee}}, \bibinfo {author} {\bibfnamefont {H.}~\bibnamefont {Kim}}, \ and\
  \bibinfo {author} {\bibfnamefont {J.}~\bibnamefont {Ahn}},\ }\href {\doibase
  10.1103/PhysRevA.95.053424} {\bibfield  {journal} {\bibinfo  {journal} {Phys.
  Rev. A}\ }\textbf {\bibinfo {volume} {95}},\ \bibinfo {pages} {053424}
  (\bibinfo {year} {2017})}\BibitemShut {NoStop}%
\bibitem [{\citenamefont {Kaufman}\ \emph {et~al.}(2015)\citenamefont
  {Kaufman}, \citenamefont {Lester}, \citenamefont {Foss-Feig}, \citenamefont
  {Wall}, \citenamefont {Rey},\ and\ \citenamefont {Regal}}]{Kaufman2015}%
  \BibitemOpen
  \bibfield  {author} {\bibinfo {author} {\bibfnamefont {A.~M.}\ \bibnamefont
  {Kaufman}}, \bibinfo {author} {\bibfnamefont {B.~J.}\ \bibnamefont {Lester}},
  \bibinfo {author} {\bibfnamefont {M.}~\bibnamefont {Foss-Feig}}, \bibinfo
  {author} {\bibfnamefont {M.~L.}\ \bibnamefont {Wall}}, \bibinfo {author}
  {\bibfnamefont {A.~M.}\ \bibnamefont {Rey}}, \ and\ \bibinfo {author}
  {\bibfnamefont {C.~A.}\ \bibnamefont {Regal}},\ }\href
  {http://dx.doi.org/10.1038/nature16073} {\bibfield  {journal} {\bibinfo
  {journal} {Nature}\ }\textbf {\bibinfo {volume} {527}},\ \bibinfo {pages}
  {208} (\bibinfo {year} {2015})}\BibitemShut {NoStop}%
\bibitem [{\citenamefont {Lengwenus}\ \emph {et~al.}(2010)\citenamefont
  {Lengwenus}, \citenamefont {Kruse}, \citenamefont {Schlosser}, \citenamefont
  {Tichelmann},\ and\ \citenamefont {Birkl}}]{Lengwenus2010}%
  \BibitemOpen
  \bibfield  {author} {\bibinfo {author} {\bibfnamefont {A.}~\bibnamefont
  {Lengwenus}}, \bibinfo {author} {\bibfnamefont {J.}~\bibnamefont {Kruse}},
  \bibinfo {author} {\bibfnamefont {M.}~\bibnamefont {Schlosser}}, \bibinfo
  {author} {\bibfnamefont {S.}~\bibnamefont {Tichelmann}}, \ and\ \bibinfo
  {author} {\bibfnamefont {G.}~\bibnamefont {Birkl}},\ }\href {\doibase
  10.1103/PhysRevLett.105.170502} {\bibfield  {journal} {\bibinfo  {journal}
  {Phys. Rev. Lett.}\ }\textbf {\bibinfo {volume} {105}},\ \bibinfo {pages}
  {170502} (\bibinfo {year} {2010})}\BibitemShut {NoStop}%
\bibitem [{\citenamefont {Levine}\ \emph {et~al.}(2018)\citenamefont {Levine},
  \citenamefont {Keesling}, \citenamefont {Omran}, \citenamefont {Bernien},
  \citenamefont {Schwartz}, \citenamefont {Zibrov}, \citenamefont {Endres},
  \citenamefont {Greiner}, \citenamefont {Vuleti\ifmmode~\acute{c}\else
  \'{c}\fi{}},\ and\ \citenamefont {Lukin}}]{Levine2018}%
  \BibitemOpen
  \bibfield  {author} {\bibinfo {author} {\bibfnamefont {H.}~\bibnamefont
  {Levine}}, \bibinfo {author} {\bibfnamefont {A.}~\bibnamefont {Keesling}},
  \bibinfo {author} {\bibfnamefont {A.}~\bibnamefont {Omran}}, \bibinfo
  {author} {\bibfnamefont {H.}~\bibnamefont {Bernien}}, \bibinfo {author}
  {\bibfnamefont {S.}~\bibnamefont {Schwartz}}, \bibinfo {author}
  {\bibfnamefont {A.~S.}\ \bibnamefont {Zibrov}}, \bibinfo {author}
  {\bibfnamefont {M.}~\bibnamefont {Endres}}, \bibinfo {author} {\bibfnamefont
  {M.}~\bibnamefont {Greiner}}, \bibinfo {author} {\bibfnamefont
  {V.}~\bibnamefont {Vuleti\ifmmode~\acute{c}\else \'{c}\fi{}}}, \ and\
  \bibinfo {author} {\bibfnamefont {M.~D.}\ \bibnamefont {Lukin}},\ }\href
  {\doibase 10.1103/PhysRevLett.121.123603} {\bibfield  {journal} {\bibinfo
  {journal} {Phys. Rev. Lett.}\ }\textbf {\bibinfo {volume} {121}},\ \bibinfo
  {pages} {123603} (\bibinfo {year} {2018})}\BibitemShut {NoStop}%
\bibitem [{\citenamefont {Murmann}\ \emph {et~al.}(2015)\citenamefont
  {Murmann}, \citenamefont {Bergschneider}, \citenamefont {Klinkhamer},
  \citenamefont {Z\"urn}, \citenamefont {Lompe},\ and\ \citenamefont
  {Jochim}}]{Murmann2015}%
  \BibitemOpen
  \bibfield  {author} {\bibinfo {author} {\bibfnamefont {S.}~\bibnamefont
  {Murmann}}, \bibinfo {author} {\bibfnamefont {A.}~\bibnamefont
  {Bergschneider}}, \bibinfo {author} {\bibfnamefont {V.~M.}\ \bibnamefont
  {Klinkhamer}}, \bibinfo {author} {\bibfnamefont {G.}~\bibnamefont {Z\"urn}},
  \bibinfo {author} {\bibfnamefont {T.}~\bibnamefont {Lompe}}, \ and\ \bibinfo
  {author} {\bibfnamefont {S.}~\bibnamefont {Jochim}},\ }\href {\doibase
  10.1103/PhysRevLett.114.080402} {\bibfield  {journal} {\bibinfo  {journal}
  {Phys. Rev. Lett.}\ }\textbf {\bibinfo {volume} {114}},\ \bibinfo {pages}
  {080402} (\bibinfo {year} {2015})}\BibitemShut {NoStop}%
\bibitem [{\citenamefont {Sturm}\ \emph {et~al.}(2017)\citenamefont {Sturm},
  \citenamefont {Schlosser}, \citenamefont {Walser},\ and\ \citenamefont
  {Birkl}}]{Sturm2017}%
  \BibitemOpen
  \bibfield  {author} {\bibinfo {author} {\bibfnamefont {M.~R.}\ \bibnamefont
  {Sturm}}, \bibinfo {author} {\bibfnamefont {M.}~\bibnamefont {Schlosser}},
  \bibinfo {author} {\bibfnamefont {R.}~\bibnamefont {Walser}}, \ and\ \bibinfo
  {author} {\bibfnamefont {G.}~\bibnamefont {Birkl}},\ }\href {\doibase
  10.1103/PhysRevA.95.063625} {\bibfield  {journal} {\bibinfo  {journal} {Phys.
  Rev. A}\ }\textbf {\bibinfo {volume} {95}},\ \bibinfo {pages} {063625}
  (\bibinfo {year} {2017})}\BibitemShut {NoStop}%
\bibitem [{\citenamefont {Gissibl}\ \emph {et~al.}(2016)\citenamefont
  {Gissibl}, \citenamefont {Thiele}, \citenamefont {Herkommer},\ and\
  \citenamefont {Giessen}}]{Gissibl2016}%
  \BibitemOpen
  \bibfield  {author} {\bibinfo {author} {\bibfnamefont {T.}~\bibnamefont
  {Gissibl}}, \bibinfo {author} {\bibfnamefont {S.}~\bibnamefont {Thiele}},
  \bibinfo {author} {\bibfnamefont {A.}~\bibnamefont {Herkommer}}, \ and\
  \bibinfo {author} {\bibfnamefont {H.}~\bibnamefont {Giessen}},\ }\href
  {http://dx.doi.org/10.1038/ncomms11763} {\bibfield  {journal} {\bibinfo
  {journal} {Nat. Commun.}\ }\textbf {\bibinfo {volume} {7}},\ \bibinfo {pages}
  {11763} (\bibinfo {year} {2016})}\BibitemShut {NoStop}%
\bibitem [{\citenamefont {Wen}\ \emph {et~al.}(2013)\citenamefont {Wen},
  \citenamefont {Zhang},\ and\ \citenamefont {Xiao}}]{Wen2013}%
  \BibitemOpen
  \bibfield  {author} {\bibinfo {author} {\bibfnamefont {J.}~\bibnamefont
  {Wen}}, \bibinfo {author} {\bibfnamefont {Y.}~\bibnamefont {Zhang}}, \ and\
  \bibinfo {author} {\bibfnamefont {M.}~\bibnamefont {Xiao}},\ }\href {\doibase
  10.1364/AOP.5.000083} {\bibfield  {journal} {\bibinfo  {journal} {Adv. Opt.
  Photonics}\ }\textbf {\bibinfo {volume} {5}},\ \bibinfo {pages} {83}
  (\bibinfo {year} {2013})}\BibitemShut {NoStop}%
\bibitem [{\citenamefont {Schlosser}\ \emph {et~al.}()\citenamefont
  {Schlosser}, \citenamefont {Tichelmann}, \citenamefont {Sch{\"a}ffner},
  \citenamefont {Ohl~de Mello}, \citenamefont {Hambach},\ and\ \citenamefont
  {Birkl}}]{Schlosser2019}%
  \BibitemOpen
  \bibfield  {author} {\bibinfo {author} {\bibfnamefont {M.}~\bibnamefont
  {Schlosser}}, \bibinfo {author} {\bibfnamefont {S.}~\bibnamefont
  {Tichelmann}}, \bibinfo {author} {\bibfnamefont {D.}~\bibnamefont
  {Sch{\"a}ffner}}, \bibinfo {author} {\bibfnamefont {D.}~\bibnamefont {Ohl~de
  Mello}}, \bibinfo {author} {\bibfnamefont {M.}~\bibnamefont {Hambach}}, \
  and\ \bibinfo {author} {\bibfnamefont {G.}~\bibnamefont {Birkl}},\
  }\href@noop {} {\bibinfo  {journal} {arXiv preprint arXiv:1902.05424}\
  }\BibitemShut {NoStop}%
\end{thebibliography}%
\end{document}